\documentstyle[12pt]{article}
\begin{document}

\def\ds{\displaystyle}
\def\beq{\begin{equation}}
\def\eeq{\end{equation}}
\def\bea{\begin{eqnarray}}
\def\eea{\end{eqnarray}}
\def\beeq{\begin{eqnarray}}
\def\eeeq{\end{eqnarray}}
\def\ve{\vert}
\def\vel{\left|}
\def\ver{\right|}
\def\nnb{\nonumber}
\def\ga{\left(}
\def\dr{\right)}
\def\aga{\left\{}
\def\adr{\right\}}
\def\lla{\left<}
\def\rra{\right>}
\def\rar{\rightarrow}
\def\nnb{\nonumber}
\def\la{\langle}
\def\ra{\rangle}
\def\ba{\begin{array}}
\def\ea{\end{array}}
\def\tr{\mbox{Tr}}
\def\ssp{{\Sigma^{*+}}}
\def\sso{{\Sigma^{*0}}}
\def\ssm{{\Sigma^{*-}}}
\def\xis0{{\Xi^{*0}}}
\def\xism{{\Xi^{*-}}}
\def\qs{\la \bar s s \ra}
\def\qu{\la \bar u u \ra}
\def\qd{\la \bar d d \ra}
\def\qq{\la \bar q q \ra}
\def\gGgG{\la g^2 G^2 \ra}
\def\q{\gamma_5 \not\!q}
\def\x{\gamma_5 \not\!x}
\def\g5{\gamma_5}
\def\sb{S_Q^{cf}}
\def\sd{S_d^{be}}
\def\su{S_u^{ad}}
\def\ss{S_s^{??}}
\def\sbp{{S}_Q^{'cf}}
\def\sdp{{S}_d^{'be}}
\def\sup{{S}_u^{'ad}}
\def\ssp{{S}_s^{'??}}
\def\sig{\sigma_{\mu \nu} \gamma_5 p^\mu q^\nu}
\def\fo{f_0(\frac{s_0}{M^2})}
\def\ffi{f_1(\frac{s_0}{M^2})}
\def\fii{f_2(\frac{s_0}{M^2})}
\def\O{{\cal O}}
\def\sl{{\Sigma^0 \Lambda}}
\def\es{\!\!\! &=& \!\!\!}
\def\ar{&+& \!\!\!}
\def\ek{&-& \!\!\!}
\def\cp{&\times& \!\!\!}
\def\se{\!\!\! &\simeq& \!\!\!}
\def\kpm{&\pm& \!\!\!}
\def\kmp{&\mp& \!\!\!}
\def\arr{\!\!\!\!&+&\!\!\!}
% .........................................................

\def\simlt{\stackrel{<}{{}_\sim}}
\def\simgt{\stackrel{>}{{}_\sim}}

% .........................................................

\title{
         {\Large
                 {\bf
Penguin induced $B \rar \eta$ transition form factor in light cone QCD
                 }
         }
      }

\author{\vspace{1cm}\\
{\small T. M. Aliev \thanks
{e-mail: taliev@metu.edu.tr}\,\,,
M. Savc{\i} \thanks
{e-mail: savci@metu.edu.tr}} \\
{\small Physics Department, Middle East Technical University, 
06531 Ankara, Turkey}
                       }
\date{}

\begin{titlepage}
\maketitle
\thispagestyle{empty}

\begin{abstract}
We calculate the penguin form factor for the $B \rar \eta \ell^+ \ell^-$
decay. This form factor is calculated in light cone QCD sum rules, including
contributions from wave functions up to twist--4, as well as mass
corrections of the light $\eta$ meson.
\end{abstract}

%\vspace{1cm}
~~~PACS numbers: 13.20.He, 13.20.--v, 14.40.--n
\end{titlepage}

\section{Introduction}
Rare B meson decays, induced by flavor changing neutral current (FCNC) $b
\rar s(d)$ transition, provide potentially the stringiest testing ground for
the Standard Model (SM) at loop level. These decays are also very suitable
looking for new physics beyond the SM. Among all decays of B mesons, the
semileptonic decays receive special attention, since their study offer one
of the most efficient ways in determination of the
Cabibbo--Kobayashi--Maskawa (CKM) matrix elements. From experimental side,
there scheduled an impressive program for study of both inclusive and
exclusive B--decays in B factories, BaBar and Belle, as well as LHC--b
machines. CLEO Collaboration \cite{R5401} has measured the branching ratios
of $B^0 \rar \pi^- \ell^+ \nu$ and $B \rar \rho^- \ell^+ \nu$ decays, from
which it is obtained that $\vel V_{ub} \ver = (3.25 \pm 
0.14 ^{+0.21}_{-0.29} \pm 0.55) \times 10^{-3}$. In extraction of $\vel
V_{ub} \ver$ from $B\rar \pi(\rho) \ell \nu$ decay, main theoretical
uncertainties come from $B\rar \pi(\rho)$ transition form factors. For an
accurate calculation of the CKM matrix elements, hadronic form factors need
to be determined more reliably. 

It should be noted that the decay modes of $B \rar K \ell^+ \ell^-~(\ell =
e,\mu)$ has recently been observed with ${\cal B} (B \rar K \ell^+ \ell^-)=
(0.75^{+0.25}_{-0.21} \pm 0.09) \times 10^{-6}$ \cite{R5402} and 
$(0.78^{+0.24+0.11}_{-0.20-0.18}) \times 10^{-6}$ \cite{R5403,R5404}. At
BaBar, an excess of events over background with $2.8\sigma$ has been
observed for the $B \rar K^\ast \ell^+ \ell^-$ decay with 
${\cal B} (B \rar K^\ast \ell^+ \ell^-) = (1.68^{0.68}_{-0.58} \pm 0.28)
\times 10^{-6}$ \cite{R5404}.

In this work we calculate the penguin form factor of the $B \rar \eta
\ell^+\ell^-$ decay in light cone QCD sum rules. The form factors
induced by the vector current in $B \rar \eta \ell \nu$ decay has already
been calculated in light cone QCD sum rules in \cite{R5405}. It should be
mentioned here that $B\rar \eta$ form factors are related to the 
$B\rar \pi$ form factors through SU(3) symmetry, which are calculated in
light cone QCD sum rules in \cite{R5406}. A detailed
description of the light cone QCD sum rule and its applications can be found
in \cite{R5407,R5408}.

Interest to $B \rar \eta \ell^+\ell^-$ and $B \rar \eta^\prime \ell^+\ell^-$
has its grounds in the fact that they can give information about
$\eta$--$\eta^\prime$ mixing angle \cite{R5409,R5410}. Soon B factories will
provide much more data and therefore a more reliable determination of the
transition form factors and as a result a more precise determination of
$\vel V_{ub} \ver$ will be possible. The extraction of $\vel V_{ub} \ver$ from
the $B \rar \eta(\eta^\prime) \ell^+\ell^-$ decay would present an efficient
and complementary alternative to its determination from $B \rar \pi(\rho)
\ell^+\ell^-$ decay.

The present work is organized as follows. In section 2, we calculate the sum
rule for the penguin form factor of the $B \rar \eta \ell^+\ell^-$ decay. 
Section 3 is devoted to the numerical analysis and the conclusion.
     
\section{Light cone QCD sum rules for the penguin form factors in 
$B \rar \eta$ transition}

The penguin form factor of the $B_d \rar \eta$ transition is defined as
\bea
\label{e5401}
\lla \eta(p) \vel \bar{d} \sigma_{\mu\nu} q^\nu (1+\gamma_5) b \ver
B(p_B)\rra = 2 i \Big[p_\mu q^2 - q_\mu (pq)\Big]
\frac{f_T}{m_B+m_\eta}~.
\eea
The starting point for the calculation of the form factor $f_T$ in Eq.
(\ref{e5401}) is the following correlator function:
\bea
\label{e5402}
\Pi_\mu(p,q) \es i \int d^4x e^{iqx} \lla \eta(p) \vel {\cal T} \Big\{
 \bar{q}\sigma_{\mu\nu} q^\nu (1+\gamma_5) b(x) \bar{b}(0) i (1-\gamma_5) q
\Big\} \ver 0 \rra \nnb \\
\es i \Pi^T [p_\mu q^2 - (pq) q_\mu ]~,
\eea
which is calculated in an expansion around the light cone $x^2=0$. The main
reason for choosing the chiral $\bar{b} i (1-\gamma_5) q$ current instead of
the $\bar{b} i \gamma_5 q$ current which has been used in the calculation of
the $B \rar \pi$ form factor \cite{R5408}, is because twist--3 wave
functions do not contribute for this choice, which are the main inputs of
the light cone QCD sum rules and which bring about the main uncertainty to
the results \cite{R5411}. 

Following the general idea QCD sum rules to obtain the penguin form factor
is by matching the representation of the correlator function in hadronic and
quark--gluon languages. Let us first consider the hadronic representation of
the correlator function. By inserting a complete set of states with the same
quantum numbers of the B meson between the
currents in the correlator, and singling out the pole term of the lowest
pseudoscalar B meson, we get
\bea
\label{e5403}
\Pi_\mu(p,q) \es \frac{\lla \eta\vel \bar{q}\sigma_{\mu\nu} q^\nu (1+\gamma_5)
\ver B \rra \lla B \vel \bar{b} i (1-\gamma_5) q \ver 0 \rra}
{m_B^2-(p+q)^2} \nnb \\
\ar \sum_h \frac{\lla \eta\vel \bar{q}\sigma_{\mu\nu} q^\nu
(1+\gamma_5) \ver h \rra \lla h \vel \bar{b} i (1-\gamma_5) q \ver 0 \rra}
{m_h^2-(p+q)^2}~, \nnb \\
\es i \Pi^T [p_\mu q^2 - (pq) q_\mu]~,
\eea
where the sum in Eq. (\ref{e5403}) describes the contributions of the higher
states and continuum. For the invariant amplitude $\Pi^T$ one can write a
general dispersion relation in the B meson momentum squared $(p+q)^2$ as
\bea
\label{e5404}
\Pi^T\Big(q^2,(p+q)^2\Big) = \int ds \frac{\rho(s)}{s-(p+q)^2}~.
\eea
The spectral density corresponding to (\ref{e5403}), is
\bea
\label{e5405}
\rho(s) = 2 \frac{f_T^\eta (q^2)}{m_B+m_\eta}\, \frac{m_B^2 f_B}{m_b} \, 
\delta (s-m_B^2) + \rho^h (s)~,
\eea
where we have used the definition
\bea
\lla B \vel \bar{b} i \gamma_5 q \ver 0 \rra = \frac{m_B^2 f_B}{m_b}~.\nnb
\eea
The first term in Eq. (\ref{e5405}) represents the ground state B meson
contribution and $\rho^h (s)$ corresponds to the spectral density of the
higher resonances and the continuum. The spectral density $\rho^h (s)$ can
be approximated by invoking the quark--hadron duality ansatz
\bea
\label{e5406}
\rho^h (s) = \rho^{QCD} (s-s_0)~,
\eea  
where $s_0$ is the continuum threshold. As a result, the hadronic
representation of the invariant amplitude $\Pi^T$ takes the following form
\bea
\label{e5407}
\Pi^T = 2 \frac{f^\eta_T(q^2) m_B^2 f_B}{(m_B+m_\eta) m_b [m_B^2-(p+q)^2]}
+ \int_{s_0}^\infty ds \frac{\rho^{QCD}(s)}{s-(p+q)^2} + \mbox{\rm
subtractions}~.
\eea

In order to obtain the sum rule for $f_T^\eta(q^2)$, we proceed to 
calculate of the correlator function from QCD side. This can be done by
using the light cone OPE method. For this purpose, we work in the large
space--like momentum regions $(p+q)^2 - m_b^2 \ll 0$ for the $b\bar{q}$ 
channel and $q^2 \ll m_b^2 - {\cal O}\mbox{\rm (few $GeV^2$)}$ for the
momentum transfer, which correspond to the small light cone distance 
$x \approx 0$ and are required by the validity of the OPE. 
After contracting $b$ quark field, we get
\bea
\label{e5408}
\Pi_\mu(p,q) = i \int d^4x e^{iqx} \lla \eta(p) \vel \bar{q}(x)
\sigma_{\mu\nu} q^\nu (1+\gamma_5) {\cal S}^b(x,0) i (1-\gamma_5)
q \ver 0 \rra~,
\eea
where ${\cal S}^b(x,0)$ is the full quark propagator. In presence of the 
background gluon field, its explicit expression can be written as
\bea
\label{e5409}
\lefteqn{
\lla 0 \vel {\cal T} \{ b(x) \bar{b}(x)\} \ver 0 \rra =
i \int \frac{d^4k}{(2 \pi)^4} \, e^{-ikx} 
\frac{\not\!{k} +m_b}{k^2-m_b^2}} \nnb \\  
\ek i g_s \int \frac{d^4k}{(2 \pi)^4} \, e^{-ikx} \int_0^1 du \Bigg[ 
\frac{1}{2} \, \frac{\not\!{k} +m_b}{(k^2-m_b^2)^2} \, G^{\alpha\beta}(ux)
\sigma_{\alpha\beta} - \frac{1}{k^2-m_b^2} \, u x_\alpha G^{\alpha\beta}(ux)
\gamma_\beta \Bigg]~,
\eea
where the first term on the right hand side corresponds to the free quark
propagator, $G^{\alpha\beta}$ is the gluonic field strength and $g_s$ id the
strong coupling constant. We see from Eqs. (\ref{e5408}) and (\ref{e5409})
that, in order to calculate the theoretical part of the correlator, the
matrix elements of the nonlocal operators between $\eta$ meson and vacuum
states are needed.

Here we would like to remark that in the following calculation
$\eta$--$\eta^\prime$ mixing will be neglected, since in octet--singlet
basis this angle is about $\theta \approx 10^0$ \cite{R5412}.     
Hence, in the above--mentioned basis, the interpolating current for 
$\eta$ meson is chosen as the SU(3) octet axial--vector current 
\bea
\label{e5410}
J_\mu = \frac{1}{\sqrt{6}} \Big( \bar{u} \gamma_\mu \gamma_5 u + 
\bar{d} \gamma_\mu \gamma_5 d - \bar{s} \gamma_\mu \gamma_5 s \Big)~.
\eea
In order to simplify the notation we will use $\bar{q} \Gamma q$ to denote
\bea 
J_\mu = \frac{1}{\sqrt{6}} \Big( \bar{u} \Gamma_\mu u + 
\bar{d} \Gamma_\mu d -\bar{s} \Gamma_\mu s \Big) \nnb~,
\eea
and introduce $F_\eta=f_\eta/\sqrt{6}$. Here, $f_\eta$ is the leptonic
decay constant of $\eta$ meson and is to be determined from the relation
\bea
\label{e5411}
\lla 0 \vel \bar{q} \gamma_\mu \gamma_5 q \ver \eta^{(\rho)} \rra = i
f_\eta p_\mu~.
\eea
It is easy to see from Eqs. (\ref{e5408}) and (\ref{e5409}) that the terms
containing even number of Dirac matrices do not give any contribution.
Remaining matrix elements can be parametrized in terms of $\eta$ meson
functions up to twist--4 defined as
\bea
\label{e5412}
\lefteqn{
\lla \eta (p) \vel \bar{q}(x) \gamma_\mu \gamma_5 q(0) \ver 0 \rra =
-i f_\eta p_\mu \int_0^1 du e^{iupx} \left[ \varphi_\eta(u) +
\frac{1}{16} m_\eta^2 x^2 A(u) \right]} \nnb \\
\ek \frac{i}{2} f_\eta m_\eta^2 \frac{x_\mu}{px} 
\int_0^1 du  e^{iupx} B(u)~, \\ \nnb \\
\label{e5413}
\lefteqn{
\lla \eta (p) \vel \bar{q}(x) \gamma_\mu \gamma_5 
g_s G_{\alpha\beta}(ux) q(0) \ver 0 \rra =  
f_\eta m_\eta^2 \left[p_\beta \ga g_{\alpha\mu}- \frac{x_\alpha p_\mu}
{px} \dr -  p_\alpha \ga g_{\beta\mu}- \frac{x_\beta p_\mu}
{px} \dr \right] } \nnb \\
\cp \int {\cal D} \alpha_i \varphi_\perp (\alpha_i) 
e^{ipx(\alpha_1 + u \alpha_3)}
+ f_\eta m_\eta^2 \frac{p_\mu}{px} (p_\alpha x_\beta - p_\beta x_\alpha)
\int {\cal D} \alpha_i \varphi_\parallel (\alpha_i) e^{ipx(\alpha_1 + u
\alpha_3)} \\ \nnb \\
\label{e5414}
\lefteqn{
\lla \eta (p) \vel \bar{q}(x)
g_s \widetilde{G}_{\alpha\beta}(ux) \gamma_\mu q(0) \ver 0 \rra =  
i f_\eta m_\eta^2 \left[p_\beta \ga g_{\alpha\mu}- \frac{x_\alpha p_\mu}
{px} \dr -  p_\alpha \ga g_{\beta\mu}- \frac{x_\beta p_\mu}
{px} \dr \right] } \nnb \\
\cp \int {\cal D} \alpha_i \widetilde{\varphi}_\perp (\alpha_i) 
e^{ipx(\alpha_1 + u \alpha_3)}
+ i f_\eta m_\eta^2 \frac{p_\mu}{px} (p_\alpha x_\beta - p_\beta x_\alpha)
\int {\cal D} \alpha_i \widetilde{\varphi}_\parallel (\alpha_i) e^{ipx(\alpha_1 + u
\alpha_3)}~,
\eea
where
\bea
\widetilde{G}_{\mu\nu} = \frac{1}{2} \epsilon_{\mu\nu\alpha\beta}
G^{\alpha\beta}~,\mbox{\rm and}~
{\cal D} \alpha_i = d\alpha_1 d\alpha_2 d\alpha_3 \delta (1-\alpha_1
-\alpha_2 - \alpha_3)~.\nnb
\eea 
In Eqs. (\ref{e5412})--(\ref{e5414}), the function $\varphi_\eta(u)$ is the
leading twist--2, $A(u)$, $\varphi_\parallel (\alpha_i)$,
$\varphi_\perp (\alpha_i)$, $\widetilde{\varphi}_\parallel (\alpha_i)$,
and $\widetilde{\varphi}_\perp (\alpha_i)$ are all twist--4 wave functions.
Inserting Eqs. (\ref{e5412})--(\ref{e5414}) and Eq. (\ref{e5409}) into 
Eq. (\ref{e5408}) and completing integration over the variables $x$ and $k$,
we get for the invariant structure
\bea
\label{e5415}
\Pi^T \es 2 F_\eta \int_0^1 \frac{du}{m_b^2-(q+pu)^2} \Bigg\{
\varphi_\eta (u) - \frac{1}{2} m_b^2 m_\eta^2 
\frac{A(u)}{[m_b^2-(q+pu)^2]^2}\Bigg\} \nnb \\
\ek 4 F_\eta m_\eta^2 \int_0^1 du u \int {\cal D} \alpha_i 
\frac{\varphi_\parallel (\alpha_i) - 2 \widetilde{\varphi}_\perp(\alpha_i)}
{\{m_b^2-[q+p (\alpha_1+ u \alpha_3)]^2\}^2} \nnb \\
\ar 2 F_\eta m_\eta^2 \int du \int {\cal D} \alpha_i
\frac{2 \varphi_\perp (\alpha_i)- \varphi_\parallel (\alpha_i)+
2 \widetilde{\varphi}_\perp(\alpha_i)-\widetilde{\varphi}_\parallel
(\alpha_i)}{\{m_b^2-[q+p (\alpha_1+ u \alpha_3)]^2\}^2}~.
\eea
The next and the last step in obtaining the sum rule for penguin form factor
is to carry out the Borel transformation with respect to the variable
$(p+q)^2$ which enhances the ground state contribution and suppresses
contributions of the higher states and the continuum. Finally, matching this
result with the corresponding invariant amplitude that is calculated in
hadronic and quark languages, we get the sum rule. Subtraction of the
continuum  contribution is performed by using quark--hadron duality (more
about subtraction of continuum and higher state contributions in light cone
QCD can be found in \cite{R5414,R5415}). Performing Borel transformation in
Eq. (\ref{e5415}), we get for the theoretical part
\bea
\label{e5416}
\lefteqn{
\ga \Pi^T \dr^B = F_\eta \Bigg\{ 2 \int_\delta^1 \frac{du}{u}
\, \varphi_\eta (u)e^{-s(u)/M^2}  - \frac{m_b^2 m_\eta^2}{2} \int_\delta^1 
\frac{du}{u^3} \, \frac{A(u)}{M^4} \, e^{-s(u)/M^2} }\nnb \\
\ek 4 m_\eta^2 \int du u \int {\cal D} \alpha_i 
\frac{\varphi_\parallel (\alpha_i) - 2 \widetilde{\varphi}_\perp(\alpha_i)}
{M^2 k^2} \, \theta (k-\delta) e^{-s(k)/M^2} \nnb \\
\ar 2 m_\eta^2 \int du \int {\cal D} \alpha_i 
\frac{2 \varphi_\perp (\alpha_i)- \varphi_\parallel (\alpha_i)+
2 \widetilde{\varphi}_\perp(\alpha_i)-\widetilde{\varphi}_\parallel
(\alpha_i)}{M^2 k^2} \, \theta (k-\delta) e^{-s(k)/M^2}\Bigg\}~,
\eea   
where
\bea
s(u) \es \frac{m_b^2-q^2 \bar{u} + m_\eta^2 u \bar{u}}{u}~,~~~~~
s(k) = s(u \rar k)~, \nnb \\
k \es \alpha_1 + u \alpha_3~,~~~~~\bar{u} = 1-u~,~~~~~\bar{k} = 1-k~, \nnb \\
\delta \es \frac{m_\eta^2+q^2-s_0+\sqrt{(m_\eta^2+q^2-s_0)^2 + 4 m_\eta^2
(m_b^2-q^2)}}{2 m_\eta^2}~. \nnb
\eea
In the same manner, performing Borel transformation in Eq. (\ref{e5407}) and
equating it to Eq. (\ref{e5416}), we finally get the following sum rule for 
the penguin form factor
\bea
\label{e5417}
f_T^\eta (q^2) = \frac{(m_B+m_\eta)m_b}{2 m_B^2 f_B}\, e^{m_B^2/M^2} \ga
\Pi^T \dr^B~.
\eea

\section{Numerical analysis}

In this section we present the result of our numerical calculations on
penguin form factor $f_T^\eta(q^2)$. It follows from Eqs. (\ref{e5416}) and
(\ref{e5417}) that the main input parameters of the sum rule (\ref{e5417})
are the $\eta$ meson wave functions. The explicit expressions of the wave
functions $\varphi_\eta(u)$, $A(u)$, $\varphi_\parallel(\alpha_i)$,
$\varphi_\perp(\alpha_i)$, $\widetilde{\varphi}_\parallel(\alpha_i)$ and
$\widetilde{\varphi}_\perp(\alpha_i)$ are all given in \cite{R5413}. The
other necessary input parameter of the sum rule is the leptonic decay
constant $F_\eta$. As has already been noted, we will $\eta$--$\eta^\prime$
mixing. Furthermore, since $\eta$ meson is an isoscalar, we have
\bea
F_\eta^d = F_\eta^u \equiv F_\eta = \frac{f_\eta}{\sqrt{6}}~,\nnb
\eea
where for the leptonic decay constant $\eta$ meson, we quote the result of a 
recent analysis which predicts $f_\eta = 159~MeV$ \cite{R5416}. Moreover, 
the leptonic decay constant B meson is chosen to have the value 
$f_B = 160~MeV$ \cite{R5414,R5417}. 

Having all these input parameters at hand, we proceed carrying out
numerical calculations. First of all, since $M^2$ is an auxiliary Borel parameter, 
we must find a region of $M^2$ where a physically measurable quantity be
practically independent of it. The lower bound of $M^2$ is determined by the
fact that nonperturbative terms must be subdominant. The upper limit of
$M^2$ is determined by the condition that the higher states and continuum
contributions are less than, for example, $30\%$ of the total result.
Our numerical analysis shows that both conditions are satisfied in the
region $8~GeV^2 \le M^2 \le 16~GeV^2$. Moreover, it should be emphasized
that light cone QCD sum rule predictions are reliable in the region of
momentum transfer square, i.e., $q^2 \le m_b^2-2 m_b \Lambda$, where 
$\Lambda$ is a typical hadronic scale having the value $\Lambda \simeq
0.5~GeV$, which yields $q^2 \simlt 18~GeV^2$. 

In Fig. (1) we present the dependence of the form factor $f_T^\eta(q^2)$ on
the Borel parameter $M^2$ at different values of momentum transfer square,
$q^2=0~GeV^2$, $q^2=5~GeV^2$ and $q^2=10~GeV^2$, at two different choices of
the continuum threshold $s_0=35~GeV^2$ and $s_0=40~GeV^2$. We observe from
this figure that, $f_T^\eta$ seems to be practically independent of the
Borel parameter $M^2$, as $M^2$ varies in the region $8~GeV^2 \le M^2 \le
16~GeV^2$.

Having this window for $M^2$, we next study the dependence $f_T^\eta(q^2)$
on $q^2$, at three fixed values of the Borel parameter $M^2=8~GeV^2$,
$M^2=12~GeV^2$ and $M^2=16~GeV^2$, picked obviously from the above--mentioned 
working region of $M^2$, again at two fixed values of the continuum threshold,
$s_0=35~GeV^2$ and $s_0=40~GeV^2$, as before. Depicted in Fig. (2) is the
dependence of the form factor on the momentum transfer
$q^2$, which clearly demonstrates that $f_\eta^T(0) = 0.16 \pm 0.03$. As we
have noted earlier, the prediction by the light cone QCD sum rule is not
reliable in the region $q^2 \ge 18~GeV^2$. In order to extend the present
result to whole physical region, we look for some convenient parametrization
of the form factor in such a way that in the region $4 m_\ell^2 \le q^2 \le
18~GeV^2$ this parametrization coincides with the light cone QCD sum rule
prediction. The best parametrization of $f_T^\eta$ with respect to $q^2$
can be written in terms of three parameters in the following way
\bea
\label{e5418}
f_T^\eta(q^2)= \frac{f_T^\eta(0)}{1-a_F \ds\frac{q^2}{m_B^2} +
b_F \ga \ds\frac{q^2}{m_B^2}\dr^2}~.
\eea
For the values of these parameters for the penguin form factor we obtain
$a_F=1.08$ and $b_F=0.09$, where the quoted errors can be
attributed to the variation in $s_0$ and $M^2$. As has already mentioned
earlier, the form factor for the $B \rar \eta$ transition can be related to
the corresponding $B \rar \pi$ transition form factor through SU(3)
symmetry. For example, the value $f_T^\eta(q^2=0)=0.17$ is obtained using
SU(3) symmetry seems to be in quite a good agreement with our prediction of 
$f_T^\eta(q^2=0)$. 
In conclusion, we have calculated the penguin form factor for the $B \rar
\eta \ell^+ \ell^-$ decay in light cone QCD sum rule method, including
contributions of wave functions up to twist--4 and mass correction of the
$\eta$ meson.

\newpage

\section*{Figure captions}
{\bf Fig. (1)} The dependence of the form factor
$f_T^\eta$ on the Borel parameter $M^2$ at $q^2=0~GeV^2$, 
$5~GeV^2$, and $10~GeV^2$, at fixed values of the momentum threshold  
$s_0=35~GeV^2$ and $s_0=40~GeV^2$.\\ \\
{\bf Fig. (2)} The dependence of the form factor
$f_T^\eta$ on the momentum transfer $q^2$ at $M^2=8~GeV^2$, 
$12~GeV^2$, and $16~GeV^2$, at fixed values of the momentum threshold  
$s_0=35~GeV^2$ and $s_0=40~GeV^2$.

\newpage

\newpage

\begin{figure}
\vskip 1.5 cm
    \includegraphics{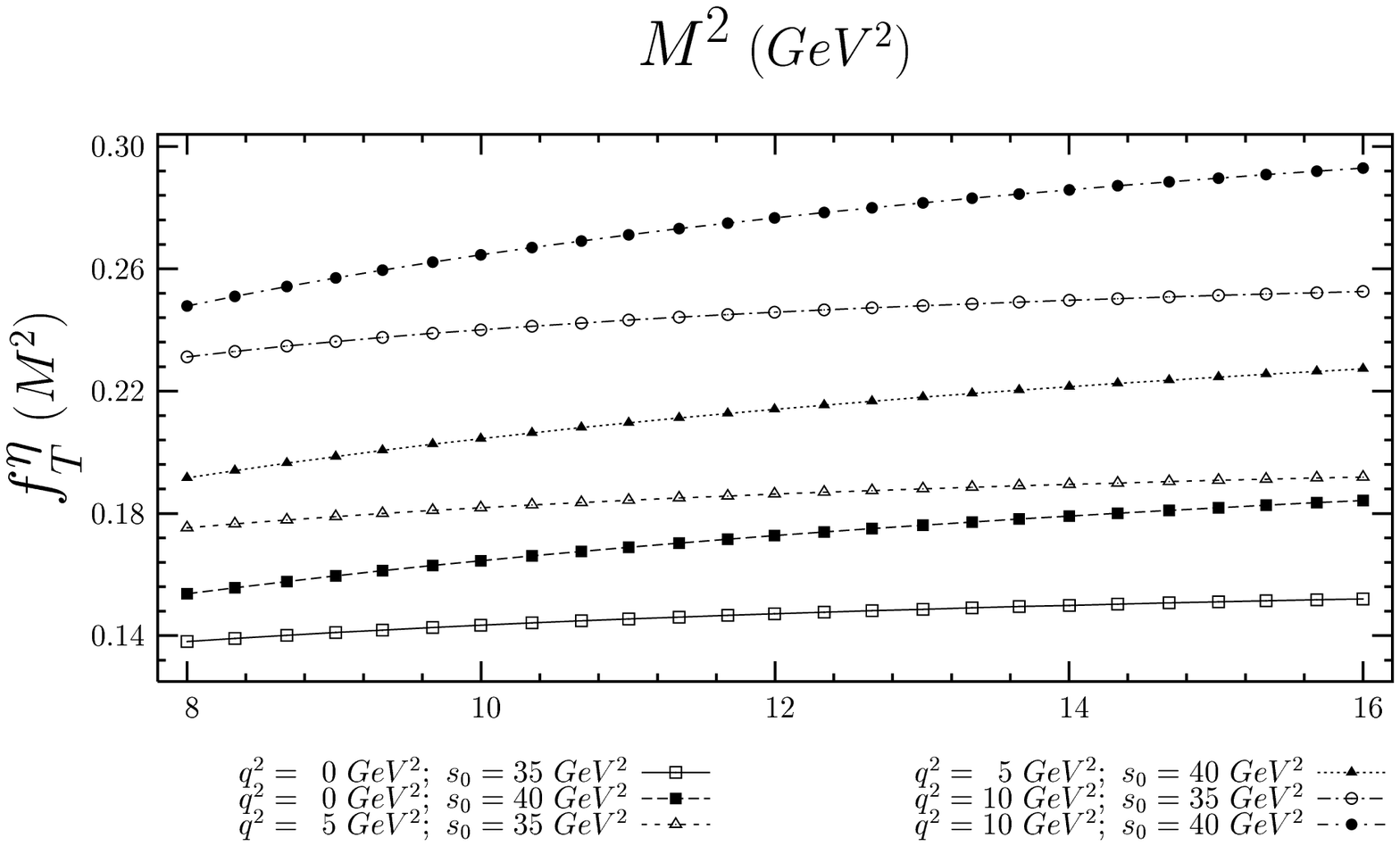}
\vskip 7.8cm
\caption{}
%\begin{center}
%{\bf Fig. 1--a}
%\end{center}
\end{figure}

\begin{figure}
\vskip 2.5 cm
    \includegraphics{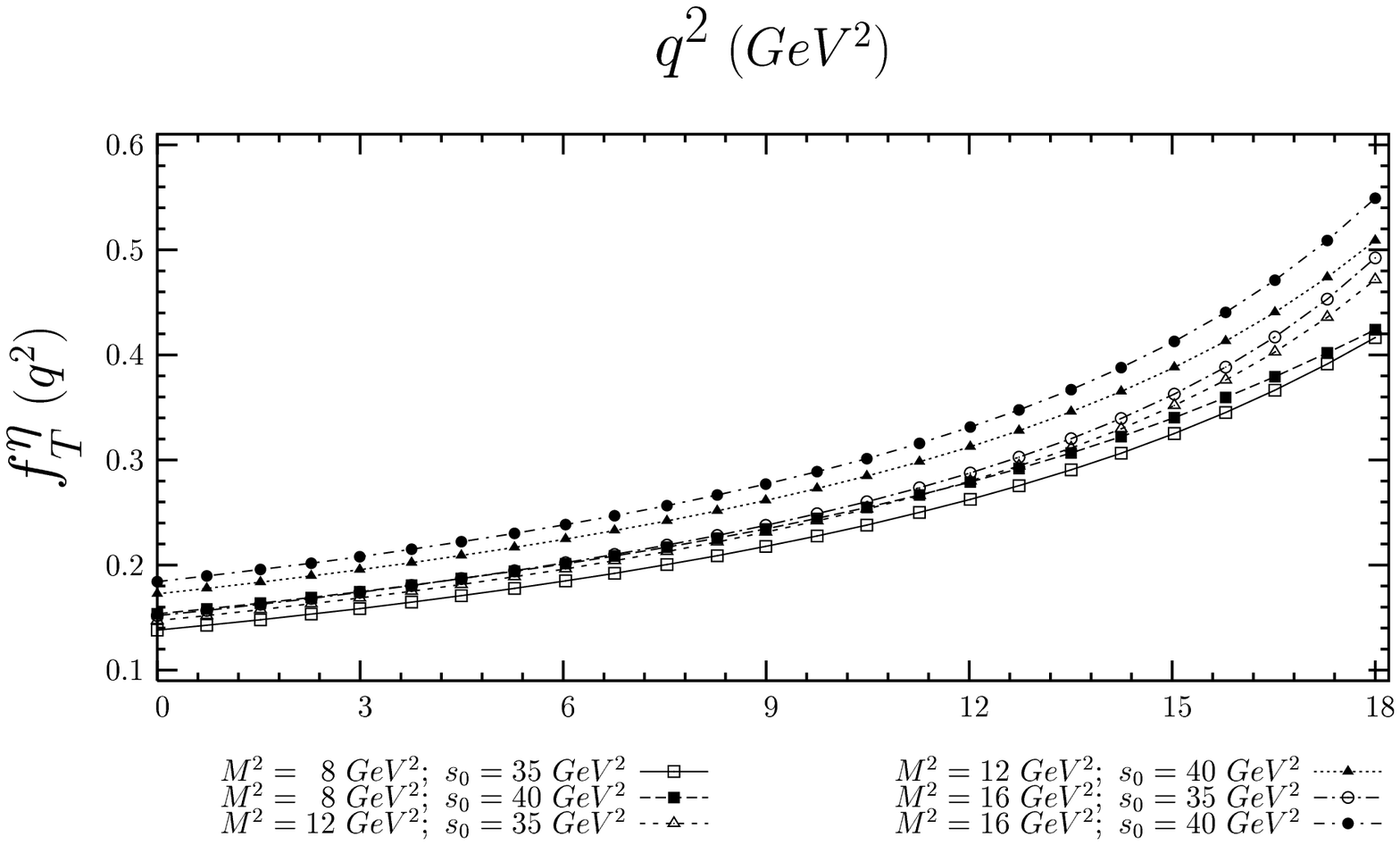}
\vskip 7.8 cm
\caption{}
%\begin{center}
%{\bf Fig. 1--b}
%\end{center}
\end{figure}

\end{document}